\documentclass{iau} 
\usepackage{graphicx}
\usepackage{natbib}

\title[Time-series analyses of Cepheid and RR Lyrae variables] 
{Time-series analyses of Cepheid and RR Lyrae variables in the wide-field variability surveys}

\author[Bhardwaj et al.]   
{A. Bhardwaj$^{1}$, 
          S. M. Kanbur$^{2}$,
           M. Marconi$^{3}$,
	   S. Das$^{4}$,\\
           E. P. Bellinger$^{5}$,
           H. P. Singh$^{4}$,
           M. Rejkuba$^{6}$
\and           C.-C. Ngeow$^{7}$} 

\affiliation{$^1$Kavli Institute for Astronomy and Astrophysics, Peking University, Yi He Yuan Lu 5, Hai Dian District, Beijing 100871, China. \\email: {\tt abhardwaj@pku.edu.cn;~anupam.bhardwajj@gmail.com} \\[\affilskip]
                $^2$State University of New York, Oswego, New York 13126, USA.\\[\affilskip]
	        $^3$INAF-Osservatorio astronomico di Capodimonte, Via Moiariello 16, 80131 Napoli, Italy.\\[\affilskip]
		$^4$Department of Physics \& Astrophysics, University of Delhi, Delhi 110007, India.\\[\affilskip]
                $^5$Stellar Astrophysics Centre, Department of Physics and Astronomy, Aarhus University, Denmark\\[\affilskip]
                $^6$European Southern Observatory, Karl-Schwarzschild-Stra\ss e 2, 85748, Garching, Germany.\\[\affilskip]
	        $^7$Graduate Institute of Astronomy, National Central University, Jhongli 32001, Taiwan.\\[\affilskip]
}

\pubyear{2018}
\volume{IAUS 347}  
\jname{Early Science with ELTs (EASE)}
\editors{N. Przybilla, K.  Pollard \& A. Calamida, eds.}
\begin{document}

\maketitle

\begin{abstract}
We discuss time-series analyses of classical Cepheid and RR Lyrae variables in the Galaxy and the Magellanic Clouds at multiple wavelengths.
We adopt the Fourier decomposition method to quantify the structural changes in the light curves of Cepheid and RR Lyrae variables. A
quantitative comparison of Cepheid Fourier parameters suggests that the canonical mass-luminosity models that lie towards the red-edge of the instability 
strip show a greater offset with respect to observations for short-period Cepheids. RR Lyrae models are consistent with observations in most
period bins. We use ensemble light curve analysis to predict the physical parameters of observed Cepheid and RR Lyrae variables using 
machine learning methods. Our preliminary results suggest that the posterior distributions of mass, luminosity, temperature and radius for 
Cepheids and RR Lyraes can be well-constrained for a given metal abundance, provided a smoother grid of models is adopted in various 
input physical parameters. 

\end{abstract}

\firstsection 
\section{Introduction}

The radially pulsating periodic variables, such as Cepheids and RR Lyraes, are crucial to our understanding of stellar evolution and pulsation. 
These variables are fundamental distance indicators and tracers of young (Cepheid) and old (RR Lyrae) stellar populations. While the Cepheid-based distance ladder
has been used extensively to estimate an increasingly accurate and precise value of the Hubble constant \citep{freedman2001, riess2018}, RR Lyrae 
can potentially calibrate a distance ladder using population II stars \citep{beaton2016} in the upcoming era of extremely large telescopes.

The time-series data for Cepheid and RR Lyrae variables are not only useful for their identification and classification but also to probe the radiation hydrodynamics of
the envelope structure of these variables \citep{smk1993}. The first quantitative study of light curve structure of these variables
dates back to \cite{slee1981}, who used Fourier amplitude and phase parameters to compare the theoretical and observed light curves. The modern theoretical pulsation models
of Cepheids and RR Lyraes \citep[][and references within]{bono1999b, marconi2015} can reproduce most observables including the light and radial velocity variations at
multiple wavelengths, thus, providing an excellent opportunity for a rigorous comparison with the multiwavelength time-series data from the massive wide-field variability surveys.

Recently, \cite{bhardwaj2015, bhardwaj2017} compiled an extensive light curve dataset for classical Cepheid variables in the Galaxy and the Magellanic Clouds to study
the variation in light curve parameters as a function of period, wavelength and metallicity. Similar work was done by \citet{das2018} for RR Lyrae variables. 
These studies employed Fourier decomposition method to quantify the variation in the light curve structure and compared their observational results with theoretical 
pulsation models. Such comparisons provide strong constraints for the various physical parameters of Cepheid and RR Lyrae variables that are used as input to the pulsation models. 
 
\begin{figure*}
\includegraphics[width=1.0\textwidth]{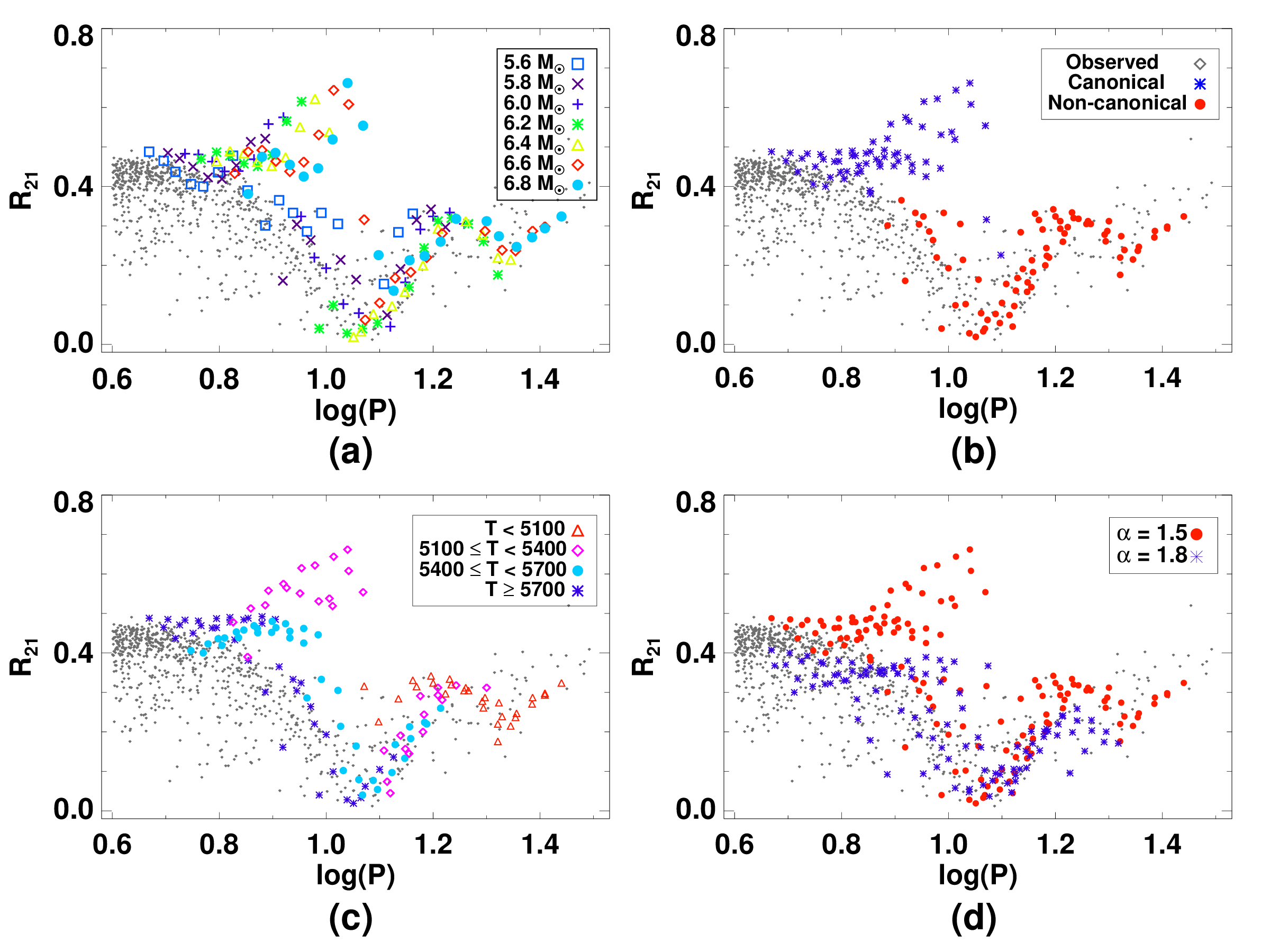}
\caption{Variation of Fourier amplitude ratio ($R_{21}$) in $I$-band for Cepheids in the LMC (small dots). The representative Cepheid models with Z=0.008 (large symbols) are
overplotted as a function of (a) stellar mass, (b) luminosity (non-canonical luminosity = canonical luminosity + 0.25 dex) (c) temperature,  and (d) mixing-length ($\alpha$).}
\label{fig:fig1}
\end{figure*}

\begin{figure*}
\includegraphics[width=1.0\textwidth]{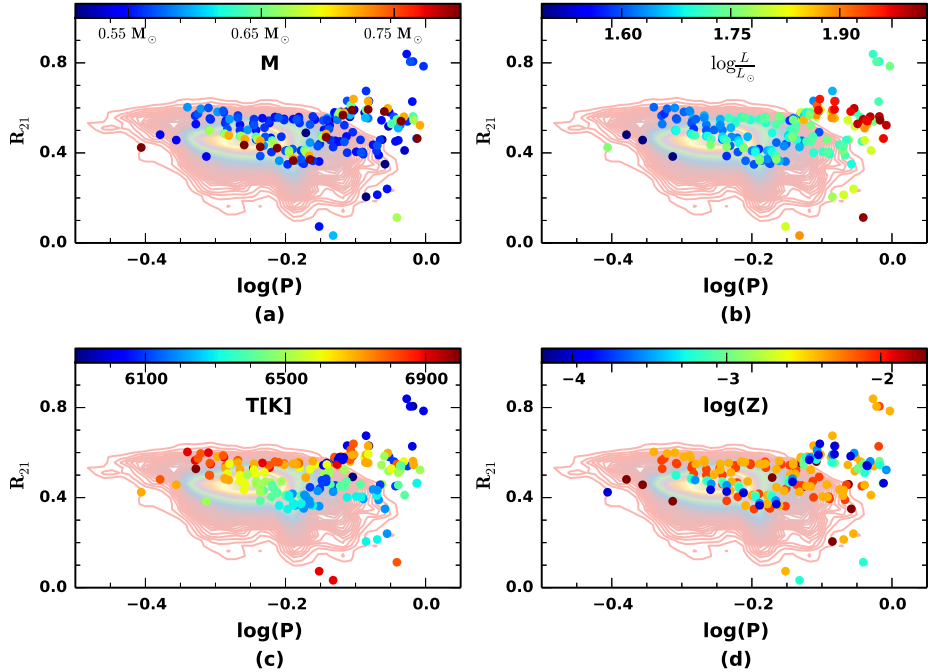}
\caption{Variation of Fourier amplitude ratio ($R_{21}$) in $I$-band for RR Lyrae in the LMC (contour maps). The RR Lyrae models (large symbols) are overplotted as a 
function of (a) stellar mass, (b) luminosity (c) temperature,  and (d) metal-abundance.}
\label{fig:fig2}
\end{figure*}

\section{Light curve analysis of Cepheid and RR Lyrae variables}

As suggested by \citet{slee1981}, Cepheid and RR Lyrae light curves can be fitted with a Fourier series in the form: $m = m_{0}+\sum_{k=1}^{N}A_{k} \sin(2 \pi k x + \phi_{k}),$
where, $x$ is the pulsation phase. They defined amplitude ratios and phase differences as: $R_{k1} = \frac{A_{k}}{A_{1}} ;~ \phi_{k1} = \phi_{k} - k\phi_{1}, \mathrm{for}~ k > 1.$ 
The observed multiband light curves of Cepheids and RR Lyraes are compiled in \citet[][]{bhardwaj2017} and \citet[][]{das2018} while the theoretical models are computed for 
a grid of physical parameters \citep{marconi2013, marconi2015}.

Fig.~\ref{fig:fig1} shows the variation in $I$-band amplitude ratio ($R_{21}$) for Cepheid variables in the LMC. 
The models representative of Cepheids in the LMC (Z=0.008, Y=0.25) are also overplotted. Note that 
canonical luminosities come from the stellar evolutionary calculations while non-canonical luminosities are brighter by 0.25 dex, to account for possible overshooting/mass-loss.
It is evident from Fig.~\ref{fig:fig1}(a \& b) that models with masses $>6\textrm{M}_\odot$ and canonical luminosities display large offsets with respect to observations
at the short-period end. Fig.~\ref{fig:fig1}(c) suggests that most of these discrepant models have low temperatures, and thus lie closer to the red-edge of the instability strip.
Fig.~\ref{fig:fig1}(d) shows that this discrepancy in amplitudes can be reduced if the convective efficiency in models is increased by increasing the mixing-length parameter. 
Fig.~\ref{fig:fig2} shows variation in $R_{21}$ for RR Lyrae in the LMC. We find that $R_{21}$ parameters from the models
with different physical parameters are consistent with the observations in most period bins.

\section{Estimating physical parameters based on light curve fitting}

The consistency of models with the observed light curves provides an opportunity to predict various physical parameters for Cepheid and RR Lyrae variables.
\citet{marconi2013} used model-fitting to a Cepheid in an eclipsing binary system and estimated mass, luminosity and temperatures that were consistent with their
precise dynamical estimates. Recently, \citet{marconi2017} used models to fit multiband light and radial velocity variations of Cepheids in the Small Magellanic Cloud 
to predict their physical parameters. We extended the model-fitting approach by adopting machine-learning methods trained on theoretical light curves \citep{bellinger2016}. 
We use various observables and Fourier coefficients to perform one-to-one comparison with models and predict physical parameters of observed variables. The preliminary results 
suggest that the posterior distributions of mass-luminosity, temperature and radius of Cepheid and RR Lyrae can be well-constrained with a smoother grid in various
physical parameters. 

\section{Conclusions}

We briefly summarized the main results from the time-series analysis of Cepheid and RR Lyrae variables in the Galaxy and the Magellanic Clouds based on Fourier decomposition methods. 
We conclude that in the era of time-domain astronomy, a global quantitative comparison of the multi-wavelength observed light curves with modern pulsation
models has the potential to provide strong constraints on physical parameters of observed Cepheid and RR Lyrae variables. However, a smooth grid of models covering the 
entire parameter space, is required for model-fitting using machine-learning methods.\\

\noindent {\it Acknowledgements:} AB acknowledges the research grant $\#11850410434$, awarded by the National Natural Science Foundation of China through a Research Fund 
for International Young Scientists.


\begin{discussion}

\discuss{Whitelock}{Very nice. Do you think there is any possibility of doing this for Miras, or are the stochastic
variations and convection going to overwhelm the physical parameters?}

\discuss{Bhardwaj}{It is possible to perform similar analysis even for Miras as long as the models are able 
to reproduce a set of observables. If AGB models can generate light variations despite all these issues, we 
can use Gaussian process methods instead of Fourier analysis to perform these comparisons, as Miras are not strictly periodic variables.}

\discuss{Anderson}{Have you already applied your methodology to different phases in a Blazhko cycle of RR Lyrae stars?
If so, what have you found?}

\discuss{Bhardwaj}{Not yet. We have only used OGLE RR Lyrae catalogue for this preliminary analysis and they do not separate Blazhko RR Lyraes.}

\end{discussion}

\end{document}